\documentclass{elsarticle}
\usepackage{mathtools}
\title{Topology and Routing Problems: The Circular Frame}

\author[1]{Rak-Kyeong Seong \corref{cor1}\fnref{fn1}}
\author[1]{Chanho Min}%\fnref{fn2}}
\author[1]{Sang-Hoon Han}% \fnref{fn3}}
\author[1]{Jaeho Yang}% \fnref{fn4}}
\author[1]{Seungwoo Nam}% \fnref{fn5}}
\author[1]{Kyusam Oh}% \fnref{fn6}}

\cortext[cor1]{Corresponding author: rk.seong@samsung.com}
\address[1]{Samsung SDS,  AI Advanced Research Lab,
  Samsung R\&D Campus, Seocho-Gu, Seoul, South Korea}

\begin{document}

%-------------------------------------------------------------------------
\begin{abstract}
   In this work, we solve the problem of finding non-intersecting paths between points on a plane with a new approach by borrowing ideas from geometric topology, in particular, from the study of polygonal schema in mathematics. 
We use a topological transformation on the 2-dimensional planar routing environment that simplifies the routing problem into a problem of connecting points on a circle with straight line segments that do not intersect in the interior of the circle.
These points are either the points that need to be connected by non-intersecting paths or special `reference' points that parametrize the topology of the original environment prior to the transformation. 
When all the necessary points on the circle are fully connected, the transformation is reversed such that the line segments combine to become the non-intersecting paths that connect the start and end points in the original environment.
We interpret the transformed environment in which the routing problem is solved as a new data structure where any routing problem can be solved efficiently. 
We perform experiments and show that the routing time and success rate of the new routing algorithm outperforms the ones for the A*-algorithm.
\end{abstract}  
\maketitle

%-------------------------------------------------------------------------
\section{Introduction\label{sec:introduction}}

The problem of identifying non-intersecting paths that connect a set of start and end points on a plane has been one of the oldest problems in computational geometry and graph theory. 
Dijkstra's algorithm \cite{dijkstra1959note} and the A*-algorithm \cite{hart1968formal} are examples of graph traversal algorithms that solve the problem of identifying shortest routes that connect start and end points on a plane. 
For our problem, we are not necessarily interested in finding the shortest or most `optimal' paths that connect a set of start and end points, but rather that all start and end points are completely connected without paths intersecting with each other. 
As expected, the problem becomes increasingly harder when the number of points increases.
By noting that the complexity of solving such problems with conventional algorithms from graph theory becomes increasingly more difficult \cite{sanders2007engineering, delling2009engineering} and their applications more varied \cite{zhang2005dynamic, nagib2010network, selamat2011fast}, we aim to solve the general routing and the related design problem from a different angle.

%%%%%
\begin{figure}[htb]
  \centering
  % <left> <lower> <right> <upper>
  \includegraphics[trim={0cm 18cm 5cm 0cm}, width=0.8\linewidth]{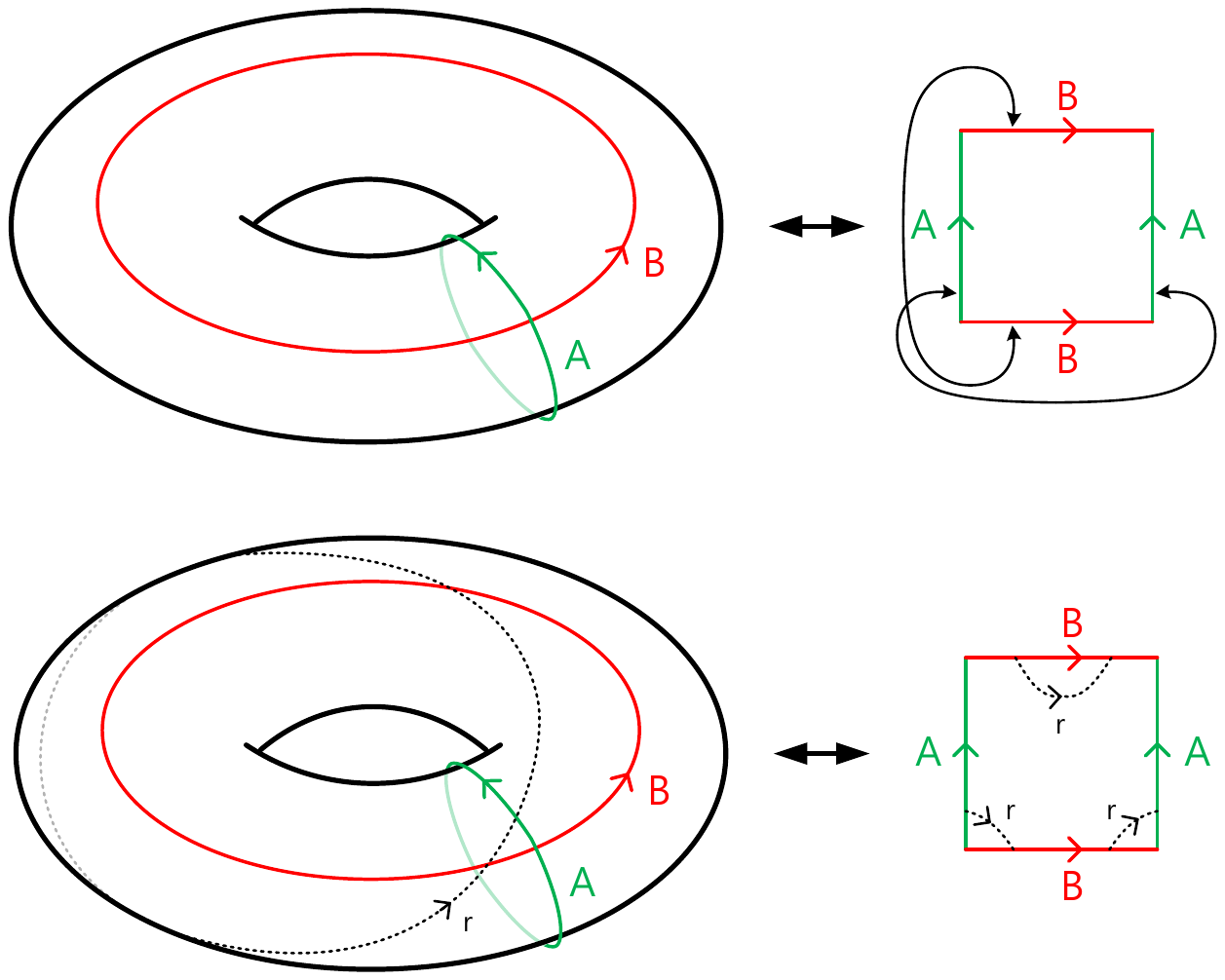}
  \caption{\label{fig:figure_torus} A 2-torus and its corresponding unit cell, which is a rectangle with its opposite sides identified with each other. A closed path $r$ on the 2-torus can be represented as a path on the unit cell, and intersections of $r$ with the boundary of the unit cell indicate the direction of $r$. The winding number of $r$ is $(n_A,n_B)=(+1,0)$.}
\end{figure}
%%%%%

In this work, we make use of geometrical topology in order to solve the problem of connecting a set of points with non-intersecting paths. 
In order to do so, we use a transformation of the planar space on which the routing is taking place.
The transformation maps the routing problem to a topologically simpler space where the problem can be solved more straightforwardly. 
Because the transformation is reversible, after all necessary start and end points are connected, the space with the routing result is transformed back to its original environment.
This transformation preserves relative rather than absolute positions of the points that need to be connected and uses concepts borrowed from geometrical topology, in particular the study of compact 2-manifolds and polygonal schema \cite{brahana1921systems, fulton2013algebraic, vegter1990computational}.

%%%%%
\begin{figure}[htb]
  \centering
  % <left> <lower> <right> <upper>
  \includegraphics[trim={0.5cm 26cm 12.5cm 0cm}, width=0.8\linewidth]{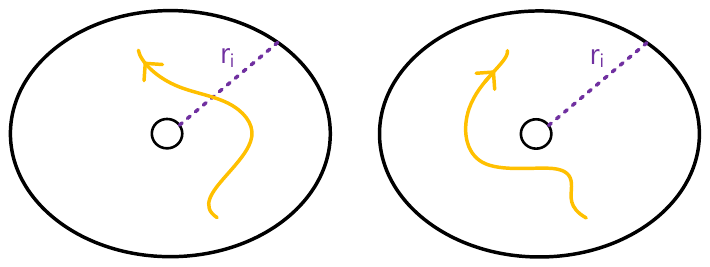}
  \caption{\label{fig:figure_orientation} A path (yellow) can pass a puncture on a bounded plane either anti-clockwise (left) or clockwise (right). The exact orientation can be determined through a reference cut or path $r_i$. In the example above, if the original path intersects with $r_i$, it runs anti-clockwise, if there is no intersection it is clockwise.}
\end{figure}
%%%%%

Polygonal schema were introduced in mathematics to study the topology of compact 2-manifolds and are particularly useful in representing the homotopy of paths on these manifolds \cite{fulton2013algebraic}.
As a result, naturally polygonal schema appeared in so-called non-crossing walk problems on compact 2-manifolds \cite{snoeyink1991topological,takahashi1993finding,papadopoulou1996k,efrat2006computing,erickson2011shortest}, where the problem is to find shortest paths in a topological space.
A combination of techniques such as the triangulation of these surfaces,  graph algorithms and the analysis of homotopy structures were used in the past to identify shortest non-crossing walks on these surfaces. 

In the following sections, we give an overview of polygonal schema and how they are used to capture the homotopy classes of paths in compact 2-manifolds. 
We then discuss how polygonal schema give rise to a much more advantageous environment in which solving routing problems become more straightforward.
Following this introduction, we describe a new algorithm using polygonal schema for solving complex routing problems.  
Following the implementation of our algorithm, we test the method's performance against the A*-algorithm and conclude with future extension and applications for computer-aided design problems.

%-------------------------------------------------------------------------
\section{Background\label{sec:background}}

A useful way of looking at compact 2-manifolds such as the Riemann surface of genus 1, i.e. a torus, is to represent it in terms of its corresponding polygonal schema. 
The 2-torus can be represented in terms of a rectangle with opposite sides of the rectangle being identified with each other as shown in Figure \ref{fig:figure_torus}.
Any such simple convex polygon together with a boundary glueing pattern is known as a polygonal schema of the represented manifold.
Using the example of the 2-torus, we learn that a rectangle with its opposite boundary sides identified with each other is topologically equivalent to a torus. 
Furthermore, it can be noted that even though a torus is 3-dimensional, it can be much more straightforwardly represented by a flat surface with an appropriate boundary identification. 
Another name for the rectangle that becomes the torus by boundary identification is the unit cell of the torus. 

When one draws a closed path on a 2-torus, it can be assigned a homotopy class or so-called winding number. 
The winding number counts the total number of turns that the path makes around a given direction. 
In case of the 2-torus, the winding number is exactly the number of times the path intersects with the boundary of the rectangle in a given direction.
Because in the case of the 2-torus, the unit cell has 2 pairs of identified boundary sections, there are two fundamental directions on the compact 2-manifold and the winding number is given by a 2-vector.
An example of a closed path and its winding number on the 2-torus is shown in Figure \ref{fig:figure_torus}.

%%%%%
\begin{figure*}[h]
  \centering
  % <left> <lower> <right> <upper>
  \includegraphics[trim={7cm 17cm 0 2cm}, width=1\linewidth]{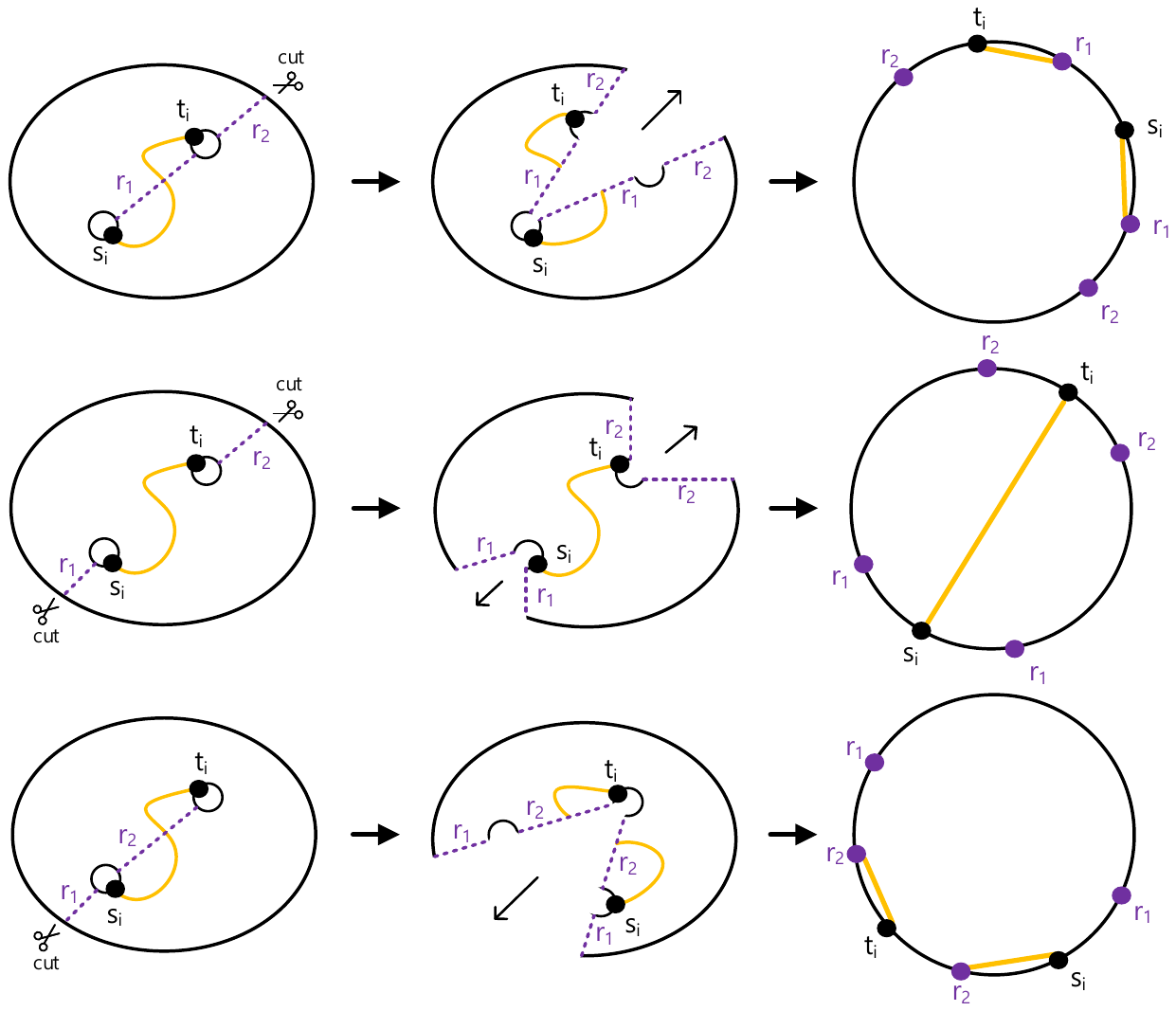}
  \caption{\label{fig:cutting} Illustration of how a pair of start and end points on a bounded plane can be transformed to the `unit cell' of the environment by cutting the plane along 3 different cut line configurations.}
\end{figure*}
%%%%%

%%%%%
\begin{figure}[h]
  \centering
  % <left> <lower> <right> <upper>
  \includegraphics[trim={1cm 23cm 11cm 0cm}, width=0.7\linewidth]{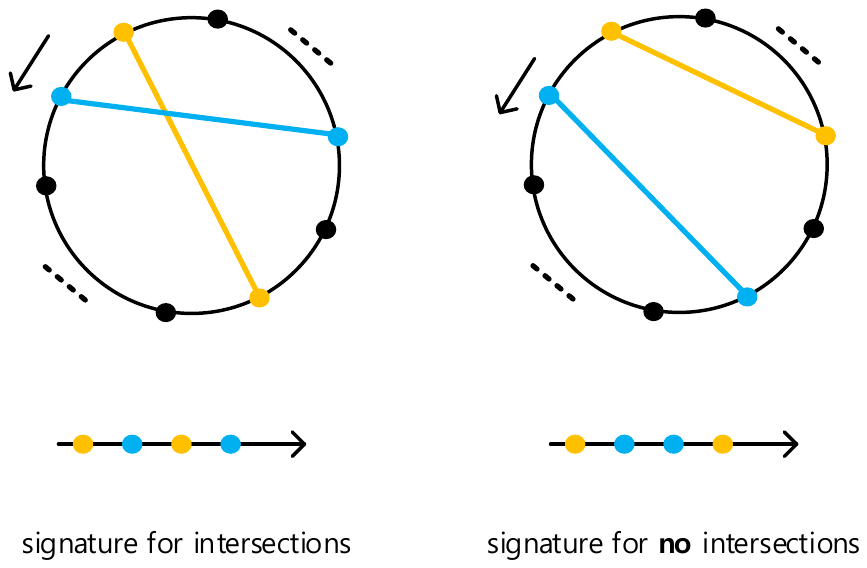}
  \caption{\label{fig:figure_intersection} A path (yellow) can pass a puncture on a bounded plane either anti-clockwise (left) or clockwise (right). The exact orientation can be determined through a reference cut or path $r_i$. If the original path intersects with $r_i$, it runs anti-clockwise, if there is no intersection it is clockwise.}
\end{figure}
%%%%%

From polygonal schemas and homotopy classes of paths, one can see that the boundary of unit cells play a crucial role in parametrizing how a path is drawn on a compact 2-manifold. 
The identified boundaries of the unit cell that form the 2-manifold cut and parametrize the direction of the path that is drawn on the 2-manifold.
The question arises whether a similar approach can be used to parametrize paths that connect a set of start and end points on a bounded plane as part of the problem that we are considering in this work.

In fact, for our problem, the topology of the environment is a punctured plane where each puncture is associated to either a start or end point of our problem. 
Given a puncture, there are two ways that a path can pass the puncture: clockwise or anti-clockwise, as illustrated in Figure \ref{fig:figure_orientation}.
In order to distinguish between these two cases, one can make use of the principle of homotopy and winding number of paths that we considered above in the context of the 2-torus.
In fact, for punctures, the principle of winding numbers exactly works the same as for polygonal schema. 
One can introduce a cut that connects to the puncture from the overall boundary of the plane such that when the path passes the puncture it either intersects with the cut or not. 
By that distinction, we can identify in each neighbourhood of a puncture the direction of a given path as illustrated in Figure \ref{fig:figure_orientation}. 
As such, for the parametrization to be complete, all punctures have to be connected to at least one cut and there is one cut for each puncture. 

One can now cut the bounded plane along the cuts to obtain the abstract polygonal schema representing the plane with punctures as illustrated in Figure \ref{fig:cutting}.
This is analogous of cutting the 2-torus along its $a$- and $b$-cycles to obtain the rectangle that forms its unit cell.  
The identified opposite sides of the unit cell rectangle represent cuts of the 2-torus and intersections of the paths with the cuts identify the direction that the path takes on the 2-torus.  
For the punctured plane the same principle holds. 
When separated along the cut lines connected to the punctures, one obtains the representative abstract unit cell and when one glues the cut lines back together one receives back the original punctured plane.

We can simplify the unit cell of the punctured plane further by mapping it into a circle.
There are points on the boundary of the circle but no points in the interior.
Edges between these points correspond to either the outer boundary of the plane or the cut lines along which we separated the punctured plane. 
We can clearly see from here that all start and end points now lie on the boundary of the circle. 
We can simplify the unit cell even further by mapping the edges that correspond to cut lines with points themselves on the boundary as shown in Figure \ref{fig:cutting}.
This amounts to pinching each side of the cut line after separation into a point. 
Under this simplification, each cut line is represented by a pair of points on the boundary of the unit cell. 
When a path enters one of these points, it comes out of the corresponding paired point and vice and versa.
This is as expected since the analogous situation is when in the punctured plane the connecting path intersects with the corresponding cut line by entering it on one side and exiting it on the other side.

For simplicity, we call the original punctured plane the `embedded frame', where absolute positions matter, and the circle with start, end and cut line points the `circular frame', where relative positions matter.
The circular frame is exactly what we identify as the abstract polygonal schema or unit cell of the original environment, the punctured plane.

A clear advantage of this representation of the original routing problem is that there are no more environmental obstacles in the interior of the routing environment.
Furthermore, the only restriction that we are worried about, the fact that paths connecting start and end point do not intersect with each other, can be checked by checking the boundary of the unit cell.
The ordering of two pairs of points on the boundary that are connected by two paths determines whether the two paths intersect in the interior or not as shown in Figure \ref{fig:figure_intersection}. 

This holographic property of the circular frame is ideally suited for introducing a data structure that encodes connectivity information as well as intersection information between a set of points.
When one goes along a fixed orientation, let's say anti-clockwise, around the unit cell of the punctured plane, each consecutive point and the indices of paths connected to it can be stored in a list that fully encapsulates the path routing.
Since each point is assigned a coordinate in the original bounded plane, the routing can be represented on the original plane by glueing together the cut lines. 

%%%%%
\begin{figure*}[h!!]
  \centering
  % <left> <lower> <right> <upper>
  \includegraphics[trim={0cm 9cm 2cm 0cm}, width=0.95\linewidth]{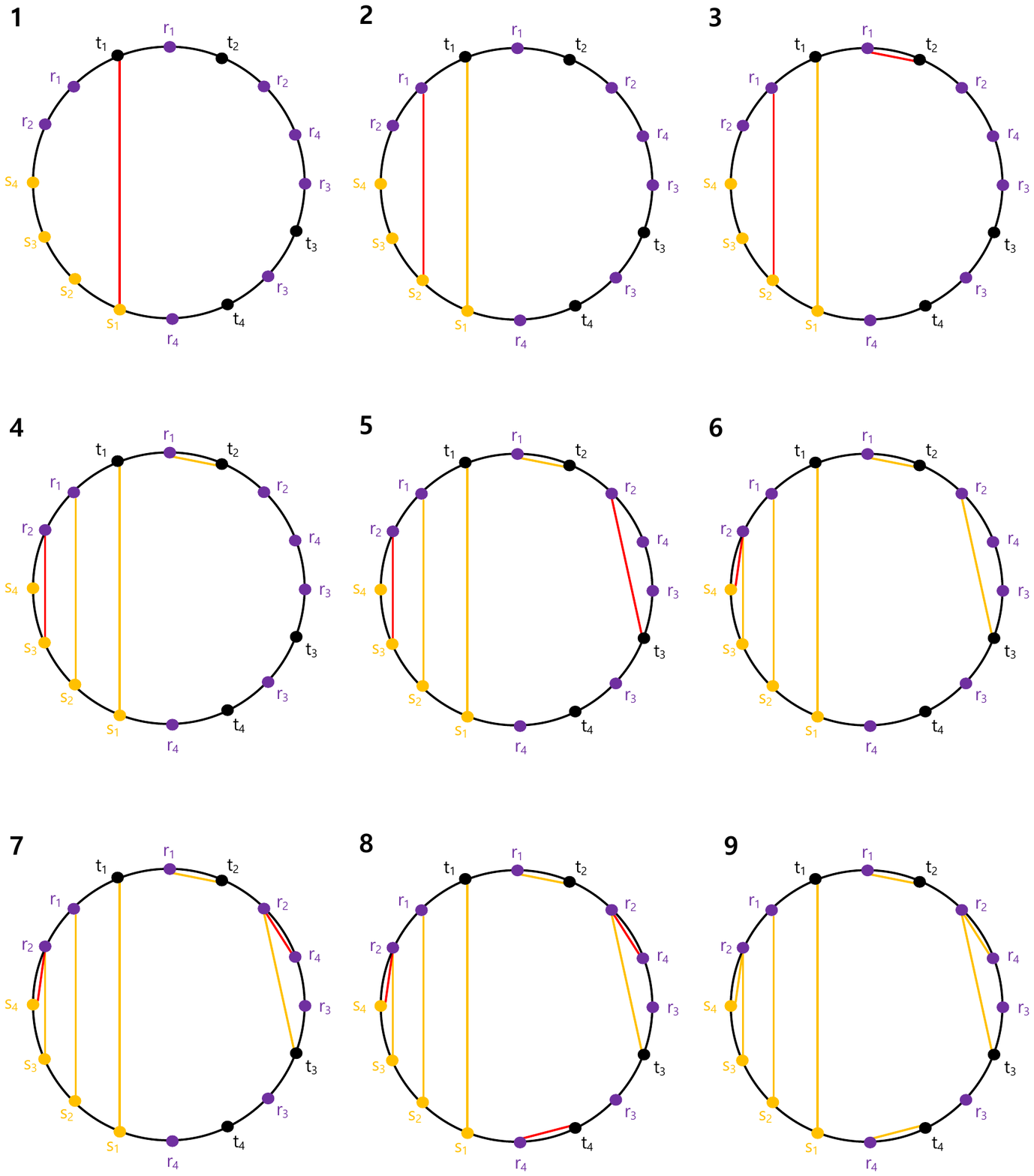}
  \caption{\label{fig:algorithm} Illustration of a simplified routing algorithm based on the circular frame. The routing steps have been following to obtain the final result.}
\end{figure*}
%%%%%

Finally, we note that connecting all associated points with non-intersecting paths is always possible. 
With the help of the circular frame this task has become a problem of connecting points on the boundary of a circle with non-intersecting line segments, without having to worry about environmental obstacles in the interior of the circle.
In the following section, we outline the full algorithm for computing routing solutions given a set of start and end points distributed in a bounded plane.

%-------------------------------------------------------------------------
\section{Method\label{sec:method}}

%%%%%
\begin{figure*}[h]
  \centering
  % <left> <lower> <right> <upper>
  \includegraphics[trim={0cm 25.5cm 7cm 0cm}, width=1\linewidth]{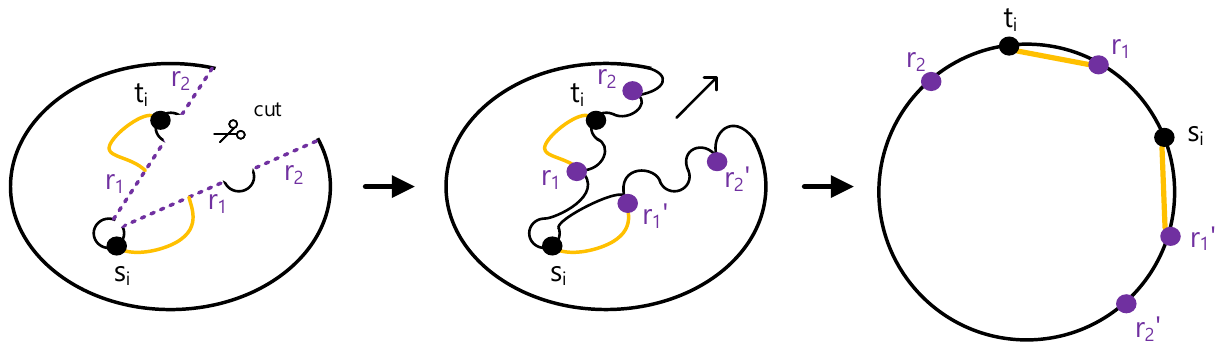}
  \caption{\label{fig:figure_pinching} The cut lines $r_i$ acting as a parametrization of the punctured plane form a tree that is connected to the outer boundary.}
\end{figure*}
%%%%%

%%%%%
\begin{figure}[h!!]
  \centering
  % <left> <lower> <right> <upper>
  \includegraphics[trim={0cm 24cm 13cm 1cm}, width=0.6\linewidth]{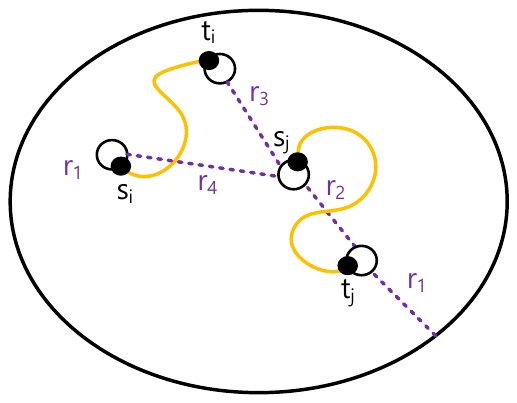}
  \caption{\label{fig:figure_tree} The cut lines $r_i$ acting as a parametrization of the punctured plane form a tree that is connected to the outer boundary.}
\end{figure}
%%%%%

For our problem, let there be set of start points $S=\{s_1,\dots,s_n\}$ and end points $T=\{t_1,\dots,t_n\}$ with pairwise identification $f:~s_i \mapsto t_i$. 
The coordinates of the vertices $s_i$ and $t_i$ are such that they are on a bounded plane $P$ with a boundary $B$.
These points are identified with punctures on the plane $P$ as illustrated in Figure \ref{fig:cutting}.

The next step is to draw a tree $R=\{r_1, \dots, r_m\}$ or a set of trees $\mathcal{R}=\{R\}$ such that the edges $r_i$ are connected with points in $S$ and $T$ or at most one point in $B$.
The possible edges are therefore $(s_i,s_j)$, $(s_i,t_j)$, $(t_i,t_j)$, $(b_i,s_j)$ and $(b_i,t_j)$, where $b_i \in B$.
Note that each tree needs to have at exactly one edge connected to a point.
Moreover, there needs to be enough trees such that every point in $S$ and $E$ is connected through a tree to the boundary $B$, as illustrated in Figure \ref{fig:figure_tree}.
The trees with their edges $r_i$ are the cuts on the plane $P$ that act as a parametrization of $P$. 
We call these reference paths $r_i$. 
Possible intersections of paths connecting $s_i$ with $t_i$ with the reference paths  determine the direction that the path is taking in $P$. 

\begin{figure*}[h]
  \centering
  % <left> <lower> <right> <upper>
  \includegraphics[trim={0cm 23.5cm 6cm 0.5cm}, width=1\linewidth]{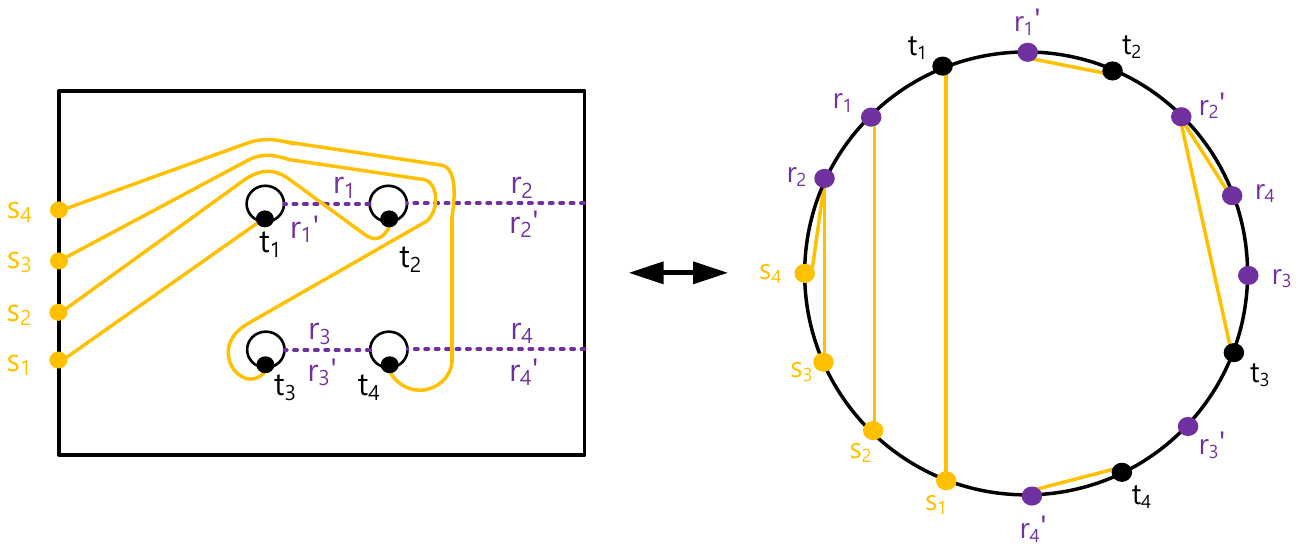}
  \caption{\label{fig:figure_routingexample} The order in which paths enter a reference path point $r_i$ is the same as the order in which they exit the paired point $r_i^\prime$.}
\end{figure*}

We cut along the reference paths $r_i$ the plane such that it becomes topologically equivalent to a circle with points on its boundary.
The points are either $s_i$, $t_i$ or correspond to points related to reference paths.
The reference paths after the cut are split into two sides  $r_i$ and $r_i^\prime$ that are pinched to points on the boundary of the circle as illustrated in Figure \ref{fig:figure_pinching}.
For simplicity, we call the circle the `circular frame' while the original setup on the plane $P$ is called the `embedded frame'. The circular frame is a simplified abstract polygonal schema representing the topology of the plane with punctures.
Figure \ref{fig:figure_routingexample} shows the correspondence between an example of the embedded frame and the circular frame of the original routing problem.

The circular frame is the primary environment where now one can execute a variety of connection algorithms. 
Let us for the purpose of this work describe a simple connection algorithm that links all start points with the corresponding end points with non-intersecting paths in the circular frame.

%%%%%
\begin{figure}[h]
  \centering
  % <left> <lower> <right> <upper>
  \includegraphics[trim={0cm 23cm 10cm 1cm}, width=1\linewidth]{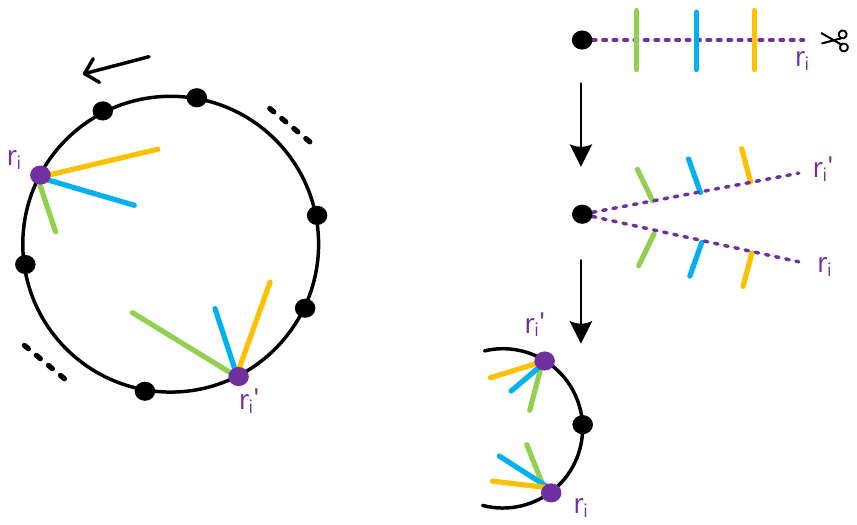}
  \caption{\label{fig:figure_ordering} The order in which paths enter a reference path point $r_i$ is the same as the order in which they exit the paired point $r_i^\prime$.}
\end{figure}
%%%%%

%%%%%
%%%%%

\begin{itemize}
\item[1.] Let there be an orientation around the boundary of the circular frame, say anti-clockwise, such that the order of appearance of starting points along the boundary under the orientation is from $s_1$ to $s_n$.

\item[2.] Starting from the first start point $s_1$, we go anti-clockwise until we reach the corresponding end point $t_1$. We draw a line connecting these two points.

\item[3.] Going to the next start point $s_i$, we go anti-clockwise until we reach:
(a) the corresponding end point $t_i$, and we draw a line connecting $t_i$ with the $s_i$ and go to step 3; or (b) a point that is already connected to a line, then we go to step 4 if that point is a reference path point $r_j$ and we draw a line connecting $s_i$ with $r_j$, or step 5 if that point is not a reference path point but a point $p_k \in ~S \cup T$; or (c) the start point $s_i$ itself, in which case we did not encounter $t_i$ on our way, and we continue with step 6.

\item[4.] Given that we are at reference path point $r_j$, we make sure that we exit $r_j$ in the same order as our path entered its pair $r_j^\prime$, and we move anti-clockwise until we reach: (a) the end point $e_i$ associated to the start point $s_i$ where we started at in step 1, and connect with a line $r_j$ and $e_i$ and go to step 3; or (b) a reference point $r_k$ that is paired with $r_k^{\prime}$ and already connected to a line, where we connect $r_j$ with $r_k$ and move to $r_k^{\prime}$ and continue with step 4 from the beginning; or (c) we reach a point $p_k$ that is not a reference point and is already connected to a line, where we move to step 5; or (c) the start point $s_i$, where we have started in step 3, then we move to step 6. Note that in (b), if there are multiple lines connected to $r_k$, we have to makes sure that the order in which our path from $r_j$ enters $r_k$ is the same as the order it exits $r_k^\prime$ as illustrated in Figure \ref{fig:figure_ordering}.

\item[5.] We are at a start or end point $p_i$ which is not associated to the original start point where we started in step 3. $p_i$ is connected to a line that pairs it at its other end with a point $p_j$. We go to $p_j$ and continue our journey anti-clockwise until we reach: (a) the end point $e_i$ corresponding to the original start point where we started in step 3, and we draw a line connecting these two points and go to step 3; (b) a reference point $r_k$ that is paired with $r_k^{\prime}$, and we move to $r_k^{\prime}$ and continue with step 3; or (c) we reach a point $p_k$ that is not a reference point but already connected to a line, where we move to step 5; or (d) the start point $s_i$ itself without having passed by its end point $t_i$, where we move to step 6. Note that in (b), if there are multiple lines connected to $r_k$, we have to makes sure the order in which our path enters $r_k$ is the same as the order it exits $r_k^\prime$ as illustrated in Figure \ref{fig:figure_ordering}.

\item[6.] We have reached $s_i$ without having encountered $t_i$. In this case, we go back clockwise along the reverse sequence of points that we have passed so far until we reach the first reference point $r_i$ and connect the previous point with $r_i$. We then continue with step 4. 

\item[7.] These steps are done until we have covered all start points $s_i \in S$. This is when all $s_i$ are connected with the corresponding $t_i$ with non-intersecting lines in the circular frame. 
\end{itemize}
Note that the above algorithm in conjunction with the circular frame representation of the routing problem guarantees that the path connecting the start and end points do not intersect with each other.
Figure \ref{fig:algorithm} illustrated the above algorithm in terms of a simple example.

The final step of the routing algorithm is to map the routing result in the circular frame back to the original environment on the bounded plane. 
We note that every start and end point on the boundary of the circular frame is assigned a coordinate in the embedded frame.
Furthermore, the pairs of reference path points $r_i$ and $r_i^\prime$ are mapped to cut lines on the embedded frame whose end points are one of the following: $(s_i,s_j)$, $(s_i,t_j)$, $(t_i,t_j)$, $(b_i,s_j)$ and $(b_i,t_j)$. 
Each of the end points have a coordinate assigned to them in the embedded frame.
When we go to the embedded frame from the circular frame, we first map the $r_i$ and $r_i^\prime$ to the separated cut lines and then sew the cut lines together to obtain a single line ending at two points with their coordinates on the embedded frame.
While doing so, we make sure that the lines entering and leaving the reference path points are distributed along the cut lines in the embedded frame in the same order as they leave and enter the points in the circular frame as illustrated in Figure \ref{fig:figure_ordering}. 

Given that no paths intersect in the interior of the circular frame, there are no paths intersecting in the embedded frame due to the reversible nature of the topological transformation that exists between these two frames.

%%%%%%%%%%%%%%%%%%%%%%%%%%%%%%%%%
\section{Experiment\label{sec:experiment}}

Let us design an experiment to compare the performance of the proposed routing algorithm based on the circular frame (CF) with the A*-algorithm (AS). 
We design a planar environment with dimensions $x_{min}=-50.0$, $x_{max}=50.0$ and $y_{min}=-50.0$, $y_{max}=500$.
The start pins $S=\{s_1,\dots,s_n\}$ and end balls $T=\{t_1,\dots,t_n\}$ are represented as circles with radius $r=0.5$.

We set $n=8$ with the coordinates of start pins $S$ set to $(50.0,28.0)$, $(50.0,20.0)$, $(50.0,12.0)$, $\dots$, $(50.0,-28.0)$. 
For $T$, the corresponding circles have coordinates that are randomly generated within the boundary of the plane in such a way that their center-to-center separation to other vertices in $T$ and $S$ is at least $d_{min}^{S,T}=1.5$ and to the boundary is $d_{min}^{B}=0.5$. 
In total, we generate $N=5000$ end ball sets $T$.

We call each generated set $(S,T)$ as a routing environment $E_{h=1\dots N}$.
For each environment $E_h$ we run both the A*-algorithm and the proposed algorithm from Section \ref{sec:method} using the circular frame. 
For the A*-algorithm we use a standard Python implementation with only moves allowed in the $x$- and $y$-directions of an integer square grid of the environment. The circular frame algorithm was also run using a Python implementation of the algorithm proposed in this work.
For each routing run, we measure the time $t_{h}$ that the algorithms take to complete the routing for all connections between paired points in $S$ and $T$.
Note that both the A*-algorithm and the circular frame algorithm were run on a standard laptop with CPU at 1.80 GHz (Intel i7-7550U) and standard memory size of 8GB.

%-------------------------------------------------------------------------
\section{Results\label{sec:results}}

\begin{figure*}[h!!]
  \centering
  %<left> <lower> <right> <upper>
  \includegraphics[trim={0cm 1cm 0cm 0cm},width=.42\linewidth]{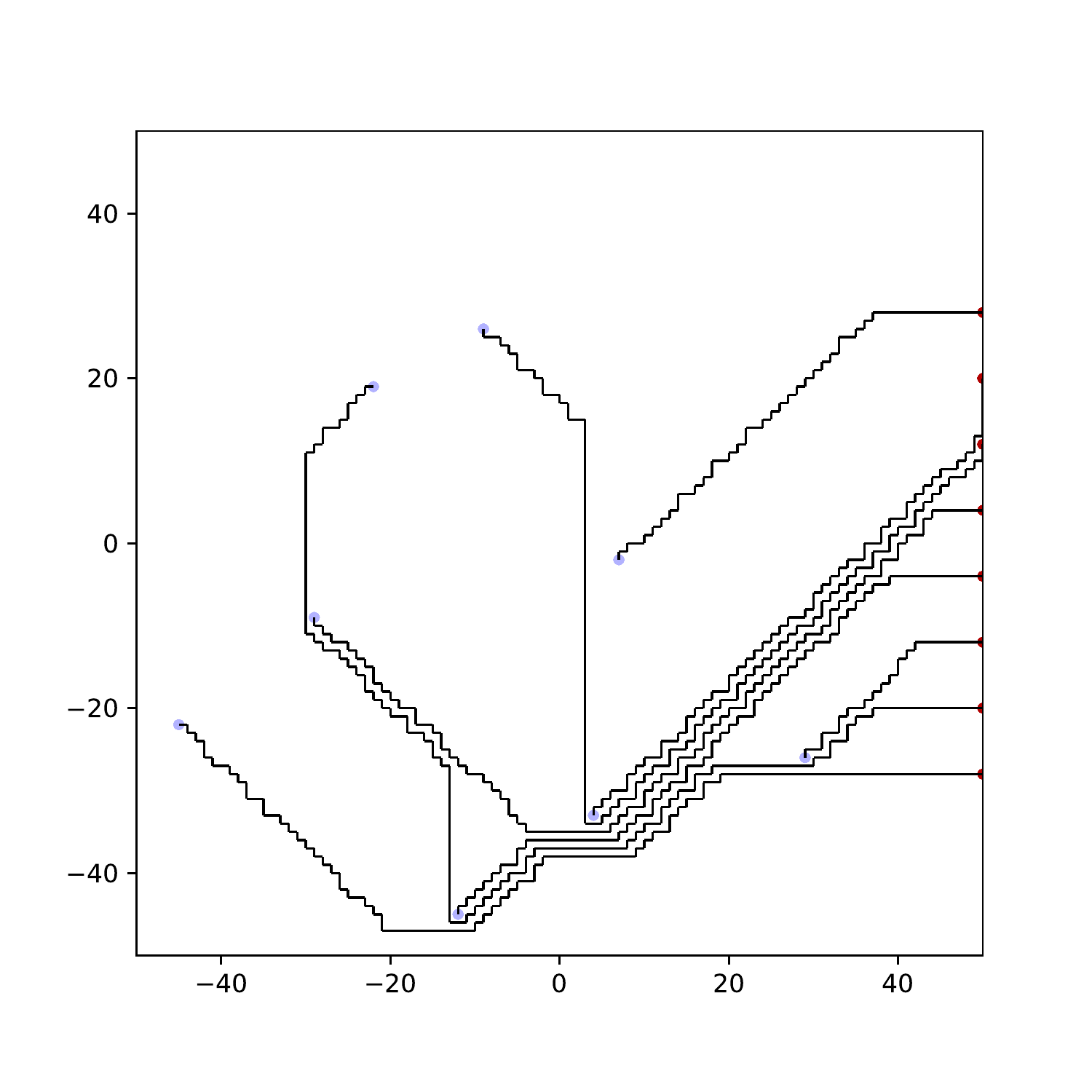}
  \includegraphics[trim={0cm 1cm 0cm 0cm},width=.42\linewidth]{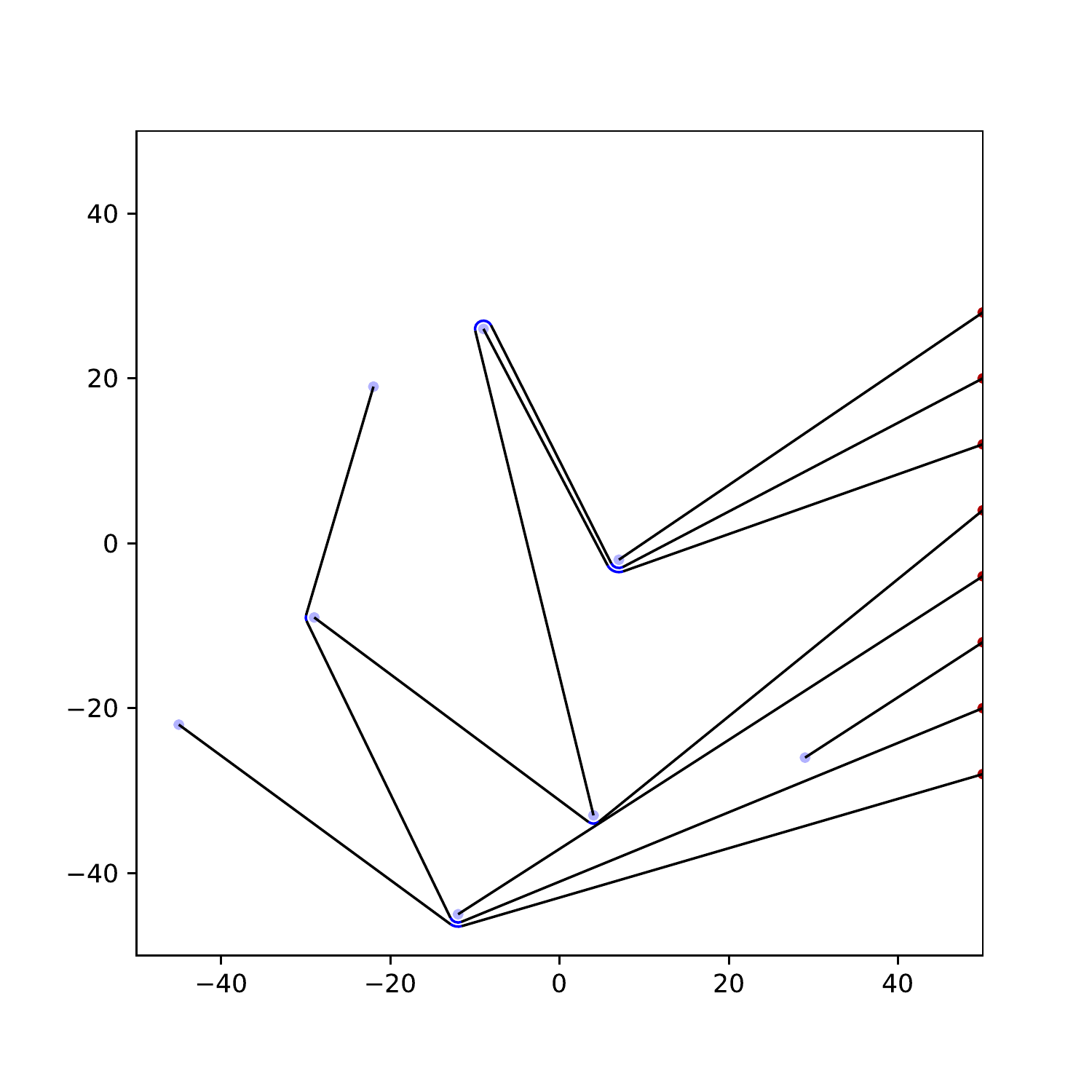}
  \\
  \includegraphics[trim={0cm 1cm 0cm 0cm},width=.42\linewidth]{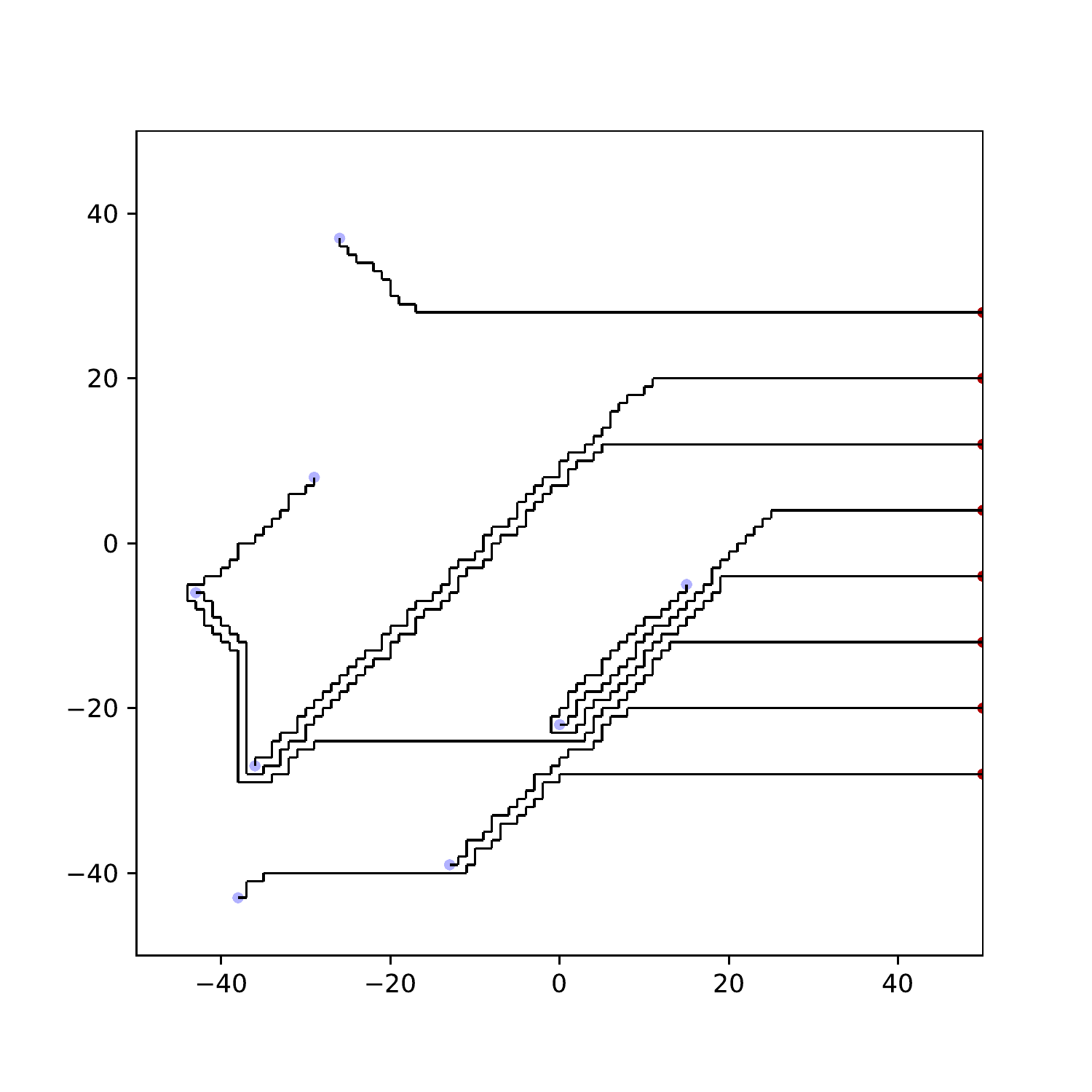}
  \includegraphics[trim={0cm 1cm 0cm 0cm},width=.42\linewidth]{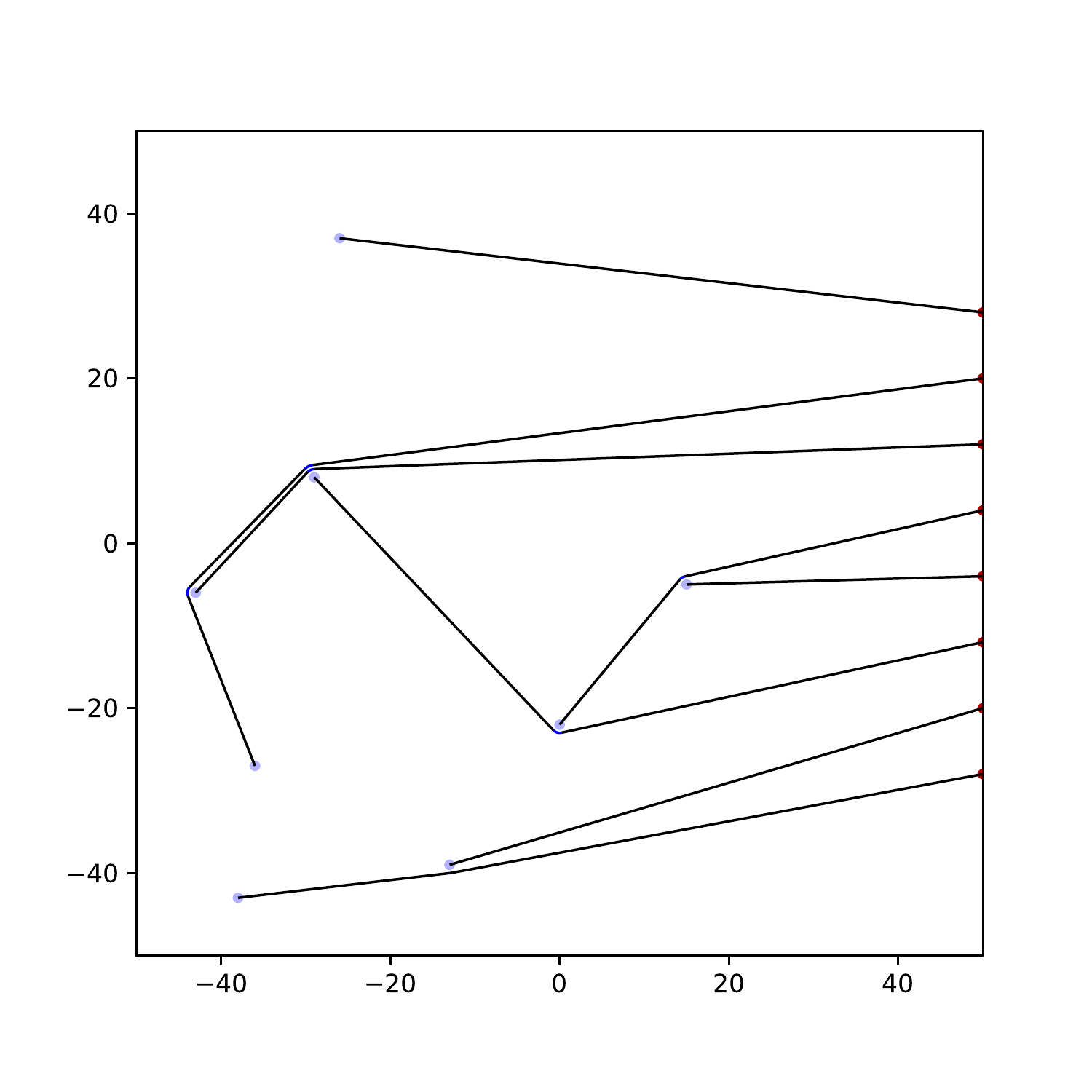}
  \\
  \includegraphics[trim={0cm 1cm 0cm 0cm},width=.42\linewidth]{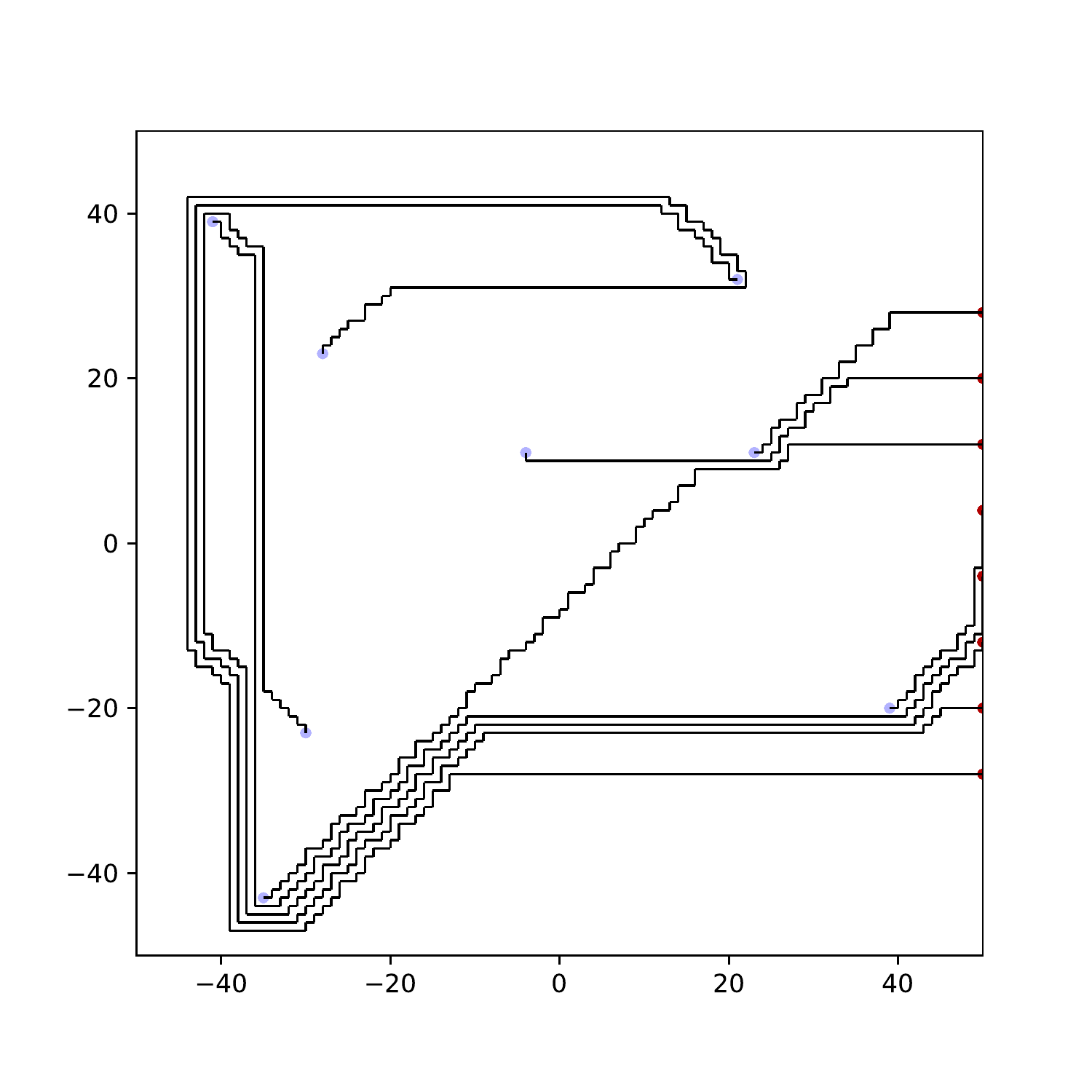}
  \includegraphics[trim={0cm 1cm 0cm 0cm},width=.42\linewidth]{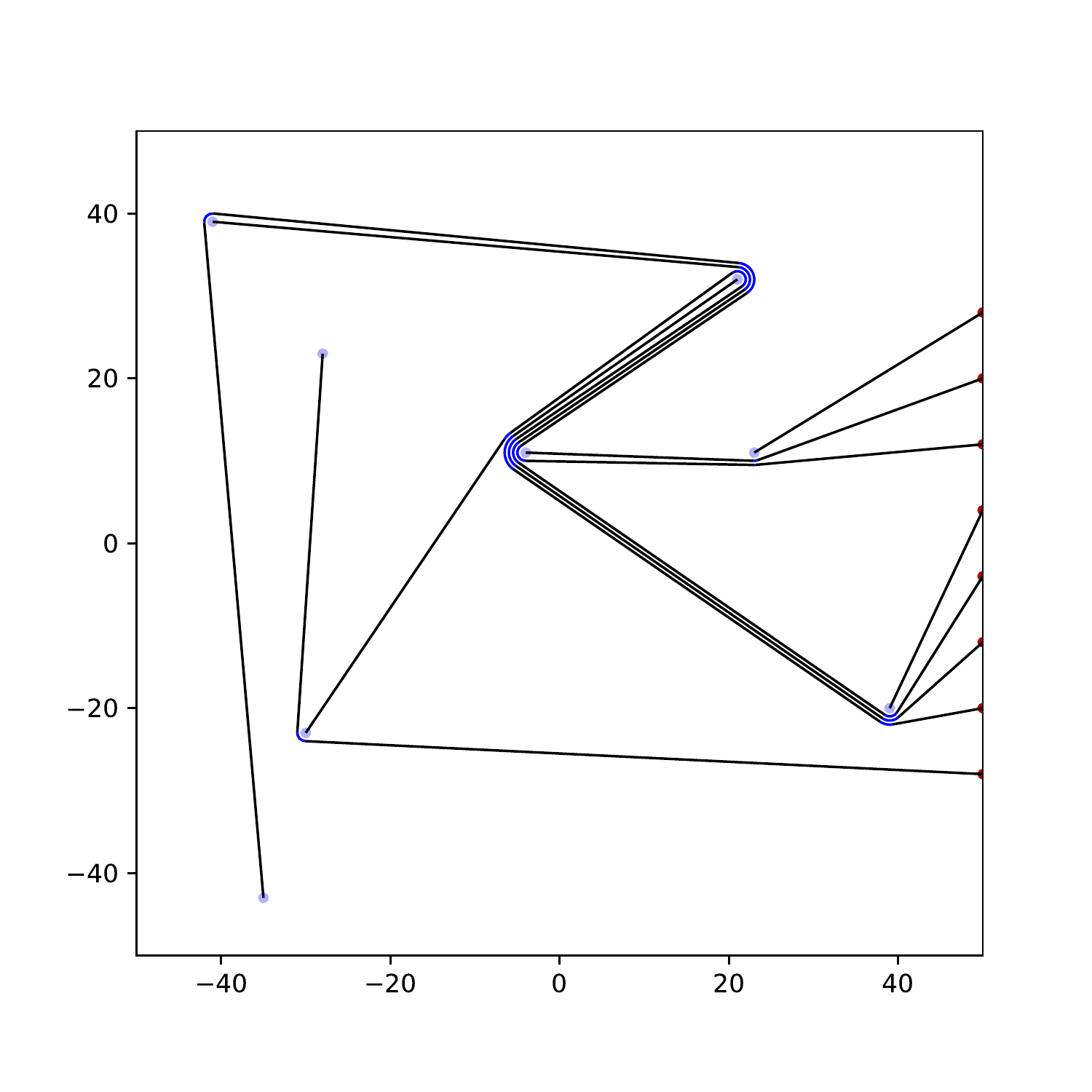}
  \caption{\label{fig:routingexamples} Three examples of randomly generated environments where both the A*-algorithm (left column) and the circular frame algorithm (right column) are applied to obtain the non-intersecting paths connecting the start and end point pairs.
  }
\end{figure*}

As part of the experiment we have conducted, we have run routings of $N=5000$ samples that were randomly generated under the restrictions outlined in Section \ref{sec:experiment}.
Out of the $N=5000$ sample environments the A*-algorithm successfully connected all start and end point pairs for only $N_{AS}=1471$ environments. 
For the remaining environments, the A*-algorithm was not able to identify paths that are non-intersecting. 
This is because the algorithm sequentially attempts to find the most optimal path between start and end point pairs, and thus the more pairs it connects, there are more obstacles to overcome for the algorithm. 
In comparison, the circular frame algorithm was able to connect all start and end point pairs with non-intersecting paths. 
This is not surprising because by construction, the circular frame algorithm easily allows us to identify non-intersecting paths with paths able to avoid themselves sequentially as outlined in Sections \ref{sec:background} and \ref{sec:method}.
Figure \ref{fig:routingexamples} shows three routing results out of the $N_{AS}=1471$ generated by the two algorithms.

\begin{table}[h]
\begin{center}
\begin{tabular}{r|l|l|l}
\hline
Algorithm & $N$ & $\overline{t}_h$ & $\sigma_{t_h}$ \\
\hline \hline
AS & $1471$ & $15.231$ & $7.288$\\
\hline
CF & 1471 & $0.046$ & $0.023$\\
\hline
CF & 5000 & $0.050$ & $0.024$ \\
\hline
\end{tabular}
\caption{The mean time $\overline{t}_h$ and corresponding standard deviation $\sigma_{t_h}$ for complete successful routing under the A*-algorithm (AS) and the circular frame algorithm (CF). Both algorithms were successful for $N=1471$ test environments, while for the remaining environments only the circular frame algorithm was successful. \label{tab:times}}
\end{center}
\end{table}

We have also measured the routing completion time for each algorithm in each generated environment $h$.
Table \ref{tab:times} shows the average routing completion times $\overline{t}_h$ for the two routing algorithms as well as the corresponding standard deviations $\sigma_{t_h}$.
One can see from the table that overall, the routing time under the implementation of the circular frame algorithm in Section \ref{sec:method} is faster than the routing time under the implementation of the A*-algorithm on the same computing machine. 

From these preliminary experimental results, we conclude that the routing performances for cases when both algorithms are successful are comparable. 
The circular frame algorithm is according to our experiment successful in connecting all start and end points with non-intersecting paths in all $N=5000$ test environments, whereas the A*-algorithm is far less reliable with $N_{AS}=1471$ successful routing results.

%%%%%
%%%%%

%-------------------------------------------------------------------------
\section{Conclusions}

In this work, we introduce methods from topology and the study of 2-manifolds to solve the problem of connecting a set of start and end points with non-intersecting paths.
Specifically, we adapted the notion of polygonal schema in such a way that it provides a topologically equivalent environment where the routing problem becomes vastly easier than the original planar environment. 
We call this new environment the circular frame.
As a circle with identified points on its boundary, the circular frame naturally represents a periodic array, which as we have shown in the above discussion, can be used to check without knowing the interior of the circle whether two lines connecting point intersect or not. 
This shows that the circular frame that we introduce here is a natural representation for routing problems where non-intersecting connections are important. 

From a computational standpoint, the routing algorithm based on the circular frame is a well-performing alternative to known algorithms such as the Dijkstra algorithm and the A*-algorithm. 
Given the circular frame representation, the new algorithm relies as mentioned before on a data structure that encodes seamlessly paths that do not intersect. 
It also enables us to identify possible intersections within the circular frame representation without having to transform it back to the original environment where coordinates are known.

From our experiments, we have shown that the circular frame routing algorithm performs similar to the A*-algorithm in terms of the morphology of the paths that make up the computed routing.
Furthermore, the success rate of completely connecting all start and end points is far higher for the circular frame algorithm than for the A*-algorithm. 
We see also that the time it takes to connect all points is faster for the new algorithm for the randomly generated data set that we use for our experiments.

In summary, we believe that using a topology-based approach to solving routing problems is an effective alternative way to conventional algorithms. 
With a focus on non-intersecting paths, our approach based on the circular frame allows us to apply this new method to various design problems in science and industry. 
In future works, we are going to expand the scope of the algorithm presented in this work, with clear outlines on which design problems we can apply our methods based on the corresponding circular frame representation.

%-------------------------------------------------------------------------
\section*{Acknowledgements}
The authors would like to thank Seungjai Min and Minsoo Kim at Samsung SDS for helpful discussions and for suggesting the problem. They are also grateful to Joung Oh Yun and Minkyu Jung at Samsung SDS for helpful guidance during the project.

%-------------------------------------------------------------------------
% bibtex
\bibliographystyle{eg-alpha-doi} 
\bibliography{mybib}       

% biblatex with biber
% \printbibliography                

%-------------------------------------------------------------------------

\end{document}